\documentclass[a4paper]{jpconf}
\usepackage{graphicx}

\DeclareRobustCommand{\rchi}{{\mathpalette\irchi\relax}}
\newcommand{\irchi}[2]{\raisebox{\depth}{$#1\chi$}} 

\newcommand{\upA}[0]{\uparrow}
\newcommand{\dwA}[0]{\downarrow}

\newcommand{\RA}[0]{\rangle}

\begin{document}
\title{Magnetic Polarizability of Virtual ($s\bar s$) and ($c\bar c$) Pairs in the Nucleon}

\author{Peter Filip}

\address{Institute of Physics, Slovak Academy of Sciences, D\'ubravsk\'a cesta 9, Bratislava 845 11, SK}

\ead{peter.filip@savba.sk}

\begin{abstract}
We suggest $^{3\!}P_{0}$ quantum state of virtual ($s\bar s$) pairs in the nucleon can be polarised by the internal 
fields permeating the volume of the nucleon (proton or neutron). Due to the quadratic Zeeman interaction, $^{3\!}P_{0}$ wavefunction of 
virtual ($q\bar q$) pairs acquires the admixture of $^{1\!}P_{10}$ quantum state in the magnetic field, which generates the 
antiparallel polarization of $s$ and $\bar s$ quarks (in the nucleon). Considering the internal magnetic fields of neutron and proton 
(originating from their measured magnetic dipole moments), we suggest the induced $s$-quark polarization in the neutron to be of 
the oposite direction compared to the proton case. We mention the influence of the internal chromo-magnetic fields on
the quantum state of  ($q\bar q$) pairs in the nucleon and we discuss also the expected behaviour of virtual ($c\bar c$) pairs.
\end{abstract}

\section{Introduction}

Internal structure of the nucleon is an important and challenging question of modern 
physics. Due to remaining uncertainities in the spin (angular momentum) structure of Proton 
(only 1/3 of the expected part of the proton spin was found \cite{SMC_spin} to be directly related to the spin 
orientations of quarks), 
the intensive experimetal efforts continue with the goal to determine the spin contribution of gluons $(\Delta G)$, and 
quark orbital angular momentum $L_q$. The spin of virtual $(s\bar s)$ pairs may also contribute 
to the total angular momentum $(J= \hbar /2)$ of proton. 

The presence of virtual $(s\bar s)$ pairs in the nucleon is naturally expected within Parton model, and the experimental data
from lepton-proton scattering as well as the observed OZI rule violations  in $\bar p p$ annihilation \cite{EllisShap} suggest
the polarization of virtual $(s\bar s)$ pairs \cite{NPA673} in the nucleon.

In this contribution we consider a polarization of $^{3\!}P_0$ ($s\bar s$) quantum state, which may occur due to
the interaction of quark magnetic (or chromo-magnetic) moments with the oriented internal electromagnetic or gluonic fields 
in the nucleon. 

\section{$^{3\!}P_0$ quantum state in external fields}

Quantum state of virtual $(s\bar s)$ component in the proton is not fully fixed by theoretical arguments, and 
several possibilities (e.g. $^{1\!}S_0$, $^{3\!}S_1$, $^{3\!}P_0$ or $^{1\!}P_1$ type of wavefunctions) exist \cite{NPA673}. Experimental data
indicate that $^{1\!}S_0$ and $^{3\!}S_1$ configurations are unlikely \cite{NPA673}, while $^{3\!}P_0$ state of ($s\bar s$) 
pair with vacuum quantum numbers $J^{PC}\! = 0^{++}$ appears to be a reasonable assumption. 

Bound state of real $(c\bar c)$ quark-antiquark pair in $^{3\!}P_0$  configuration ($0^{++}$) is observed as 
$\rchi_{c0}$ meson \cite{PDG}, and the properties of $^{3\!}P_{0}$ state in 
Positronium ($e^+e^-$)  are well understood \cite{PositroniumA}.   If we presume the response of
($e^+e^-$) and $(q\bar q)$ bound systems (in $^{3\!}P_{0}$ state) to the external field to be similar, the
existing knowledge about the Positronium $^{3\!}P_{0}$ state behavior  in external fields  \cite{PositroniumA} can be used 
to make the predictions regarding the state of virtual $(s\bar s)$ pairs in the nucleon.

\subsection{Internal magnetic fields in the nucleon}
Neutron and proton have anomalous magnetic moments  ($\mu_{n} = -1.91\mu_N$ and $\mu_p = 2.79\mu_N$) due to
their nontrivial internal structure. Using SU(6) symmetric  constituent quark $|qqq\rangle$ wavefunction, the nucleon
magnetic moment is \cite{PDG}: $\mu_N = (4\mu_{q_a} - \mu_{q_b})/3$. Constituent quark magnetic moments then are:
$\mu_u = 1.85\mu_N$ and $\mu_d = -0.97 \mu_N$. Hyperons $\Sigma^{\pm}$ and $\Xi^{0,-}$ also agree
with such a simple ansatz (within 20\% precision) and $\mu_s \approx -0.61\mu_N$ value of $s$-quark magnetic moment 
agrees well also with the magnetic moments of $\Lambda^0$ and $\Omega^-$ ($3\mu_s$) baryons \cite{PDG}. 

Magnetic moments ($\mu_p, \mu_n$) of proton and neutron
are the source of dipole magnetic field $B_{\vec\mu}$, which is generated within the interior (MIT bag volume)
of the nucleon. Compton length of the proton $(\lambda_c =  h/m_p c = 1.3^{\,}$fm) is comparable to the
proton radius (size), and one can imply a dipole field $B_{\vec\mu}$ does penetrate the interior (partonic matter) 
of the nucleon. Direction of internal magnetic field $B_{\vec\mu}$ within the proton (or neutron) volume is parallel (or antiparallel) 
to the spin orientation, and its magnitude is approximately $|B_{\vec\mu}| \approx 10^{13\,}$T = $10^{17}$Gauss. This estimate
can be obtained at semi-classical level, requiring the electric current $I$ inside the coil of radius $R_p \approx 0.82^{\,}$fm to generate
the magnetic dipole of size $\mu = \pi R_p^2 I \approx 2\mu_N$. Magnetic field inside such a theoretically
considered coil \cite{myDSPIN2013arxiv} then is: 
$B = \mu_0 I/2R_p \approx 10^{13\,}$T (here $\mu_0 = 4\pi \times 10^{-7}N/A^2$).


\subsection{Magnetic polarizability of $^{3\!}P_0$ state}
Let us consider the  wavefunction  of bound fermion-antifermion $(f\!\bar f)$ pair in $^{3\!}P_{0}$ quantum state
with $J^{PC} = 0^{++}$ quantum numbers. Using the Positronium notation \cite{Positr16x16} we have
\begin{equation}
^{3\!}P_{00} = \frac{1}{\sqrt{3}} \big (Y_{11} \rchi_{1-1} + Y_{1-1} \rchi_{11} - Y_{10} \rchi_{10}  \big ) 
\label{Pure3Po}
\end{equation}
where $Y_{lm}$ are orbital (spherical harmonics) functions, and $\rchi_{s, s_z}$ denote the spin part of $^{3\!}P_{00}$ wave function.
Now we shall assume that quantum state of the virtual $(s\bar s)$ pairs
in the nucleon has a similar (the same) quantum structure as given by Eq.(\ref{Pure3Po}).
 Using the uncoupled spin eigenstates $\rchi_{1-1} = |\!\dwA\dwA\RA$, and $\rchi_{1+1} = |\!\upA\upA\RA$,
and $\rchi_{10} = (|\!\dwA\upA\RA + |\!\upA\dwA\RA)/\sqrt{2}$, one has
\begin{equation}
^{3\!}P_{00} [s\bar s]
= \frac{1}{\sqrt{3}} \Big [
Y_{11} |\!\dwA\dwA\RA + Y_{1-1} |\!\upA\upA\RA - 
Y_{10} \frac{|\!\dwA\upA\RA + |\!\upA\dwA\RA}{\sqrt{2}} 
\Big ] 
\label{Pure3Pouu}
\end{equation}
The probability of
antiparallel $s,\bar s$ spins ($\upA\dwA$ or $\dwA\upA$) is 1/3, 
and parallel orienations $(\upA\upA)$ and $(\dwA\dwA)$ occur with the same
probability (1/3). For the net helicities of $s$ and $\bar s$ quarks in such $^{3\!}P_{00}$ state we 
have $\Delta s = s^{\upA} - s^{\dwA} = 0$ and $\Delta \bar s = \bar s^{\upA} - \bar s^{\dwA} = 0$, and $L_s + L_{\bar s} = 0$ also applies. 

It has been suggested \cite{AlbergPLB} that  component $Y_{11}|\!\dwA\dwA\RA$ in $^{3\!}P_{00}[s\bar s]$ 
wave function (\ref{Pure3Pouu})  is enhanced in the proton,
due to strong attraction of the valence $u$ and virtual $\bar s$ quarks in the pseudoscalar channel, which
results in the polarization of strange quarks in the nucleon. 
Magnetic fields can also modify the spin structure of $^{3\!}P_{00}$ state.
In the Positronium \cite{Positr16x16} case, 
$^{3\!}P_{00}$ wavefunction acquires the admixture of $^{1\!}P_{10}$ state.   
Eigenstates $^{3\!}\tilde P_{00}, ^{1\!}\tilde P_{10}, ^{3\!}\tilde P_{20}$  in 
small magnetic fields are \cite{Positr16x16}:
\begin{eqnarray}
^{3\!}\tilde P_{00}(B)  &\approx&\,\!  ^{3\!}P_{00} + \varepsilon_B \,\! ^{1\!}P_{10}/\sqrt{3} \\
^{1\!}\tilde P_{10}(B)  &\approx&\,\!  ^{1\!}P_{10} - \varepsilon_B \,\! ^{3\!}P_{00}/\sqrt{3} \, -\, \varepsilon_B' \,\! ^{3\!}P_{20} \sqrt{2/3} \\
^{3\!}\tilde P_{20}(B) &\approx&\,\!  ^{3\!}P_{20}  + \quad \!\!\! \qquad \qquad +\, \varepsilon_B' \,\! ^{1\!}P_{10} \sqrt{2/3}
\end{eqnarray} 
where admixture parameters $\varepsilon_B = \frac{2\mu_e B}{\Delta E_{10}}$ 
and $\varepsilon_B' = \frac{2\mu_e B}{\Delta E_{21}}$ depend on the magnetic field $B$, and on eigenenergy differences 
$\Delta E_{10}, \Delta E_{21}$. Strange quarks also have their magnetic moment $\mu_s$, and therefore,
a similar behavior of $^{3\!}P_{00} [s\bar s]$ wavefunction in external $B$ field can be anticipated. 

\leftline{Using $^{1\!}P_{10} = Y_{10}(|\!\dwA\upA\RA - |\!\upA\dwA\RA)/\sqrt{2}$, the
virtual $^{3\!}\tilde P_{00}$ state  of   $(s\bar s)$ pair in {\it small\,} magnetic fields 
is:}
\begin{equation}
^{3\!}\tilde P_{00}[s\bar s](B) \approx  
\frac{1}{\sqrt{3}} \Big [ Y_{11} |\!\dwA \dwA\RA + Y_{1-1}  |\!\upA \upA\RA  \Big ]
- \frac{Y_{10}}{\sqrt{3}} \Big( |\!\dwA\upA\RA\frac{1 +\varepsilon_B}{\sqrt{2}} + |\!\upA\dwA\RA\frac{1 -\varepsilon_B}{\sqrt{2}} \Big)
\label{MixingB03PoSpin}
\end{equation}
Relative imbalance of $\upA\dwA$ and $\dwA\upA$ spin configurations in the wavefunction (\ref{MixingB03PoSpin})
generates the induced magnetic moment of $^{3\!}\tilde P_{00}[s\bar s](B)$ state. Such behavior corresponds to magnetic 
polarizability \cite{Petrunkin}. 

\begin{figure}
\begin{center}
\includegraphics{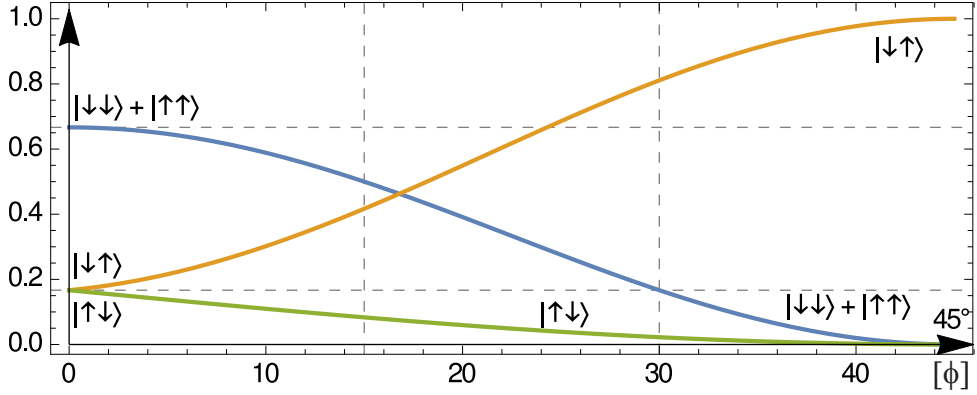}
\end{center}
\vspace{-2mm}
\caption{\label{labelF1}Probability of the spin components in $^{3\!}\tilde P_{00}$ state in the magnetic field, as a function
of the mixing angles (value $\phi\! =\! 45^\circ$ corresponds to Paschen-Back limit of very large fields: $B\rightarrow \infty$). }
\end{figure}

Energy of the Positronium $^{3\!}\tilde P_{00}$ state decreases in the magnetic field \cite{PositroniumA}, which corresponds to
$\uparrow_{\!\mu_{{e^-}}\!}\uparrow_{\!\mu_{e^+}}$ orientation of $e^-,e^+$ magnetic moments. 
In Paschen-Back limit (see Fig.$^{\,}$\ref{labelF1}), component $|^{\!}\!\dwA\upA\RA$ becomes dominant in $^{3\!}P_{00}$
state. 
Therefore, one can expect enhanced probability of  $\uparrow_{\!\mu_s\!}\uparrow_{\!\mu_{\bar s}}$ configuration $|\!\dwA\upA\RA$ 
for the virtual $(s\bar s)$ pairs in the polarized proton as well.
The behavior of $^{3\!} P_{00}$ Positronium wavefunction in the electric field is well known \cite{Positr16x16}, and our observation (analysis)
gives: No polarization of $s,\bar s$  spins occurs, if $^{3\!}\tilde P_{00}$ state is exposed to the electric fields. 


\section{Virtual $s\bar s$ and $c\bar c$ pairs in the proton and neutron}
If other effects (e.g. pseudoscalar QCD interaction \cite{AlbergPLB}) do not determine the polarization of ($q\bar q$) 
pairs in the nucleon, and the polarizability of virtual ($s\bar s$) pairs by the internal magnetic field is indeed significant, 
the following qualitative predictions 
can be inferred from Eq.(6):  

\begin{itemize}

\item Net spin polarization of $s$ quarks in proton is {\it antiparallel\,} to the proton spin.

\item Net spin polarization of $s$ quarks in the neutron is {\it parallel\,} to the spin of the neutron.

\item Polarization of $\bar s$ quarks in the nucleon favors to be opposite to the $s$ quark polarization.

\item The polarization of $c$ quarks in the nucleon is opposite to the polarization of $s$ quarks. 

\item The polarization of $\bar c$ (and $\bar u$) quarks is parallel to the polarization of $s$ quarks. 

\end{itemize}

The first two statements originate from the internal magnetic dipole  field $B_{\vec \mu}$ orientation 
relative to the spin of proton 
and neutron,
and the last two statements come from the fact, that $c$ quarks have positive charge (2/3)$e$, and thus their magnetic moment is parallel
to their spin (contrary to the case of negatively charged $s$ quarks). The third statement applies only if
$\uparrow\uparrow$ and $\downarrow\downarrow$ components of  the function (\ref{MixingB03PoSpin}) 
have the same probability (QCD interaction \cite{AlbergPLB} is inactive).


\section{Magnetic moments of virtual quarks and magnitude of the effect}
Magnetic polarizability of $(q\bar q)$ pair depends on the quark magnetic moment $|\mu_q|$. The observed
magnetic moments of baryons \cite{PDG} require $\mu_u$=1.85, $\mu_d$=-0.97, $\mu_s$=-0.61 
(in $\mu_N$ units),
for the {\it constituent} quarks $u, d, s$. This agrees with Dirac equation (demanding $\mu_q$=$\hbar Q/2 m^*_q$), 
if constituent quark masses are: $m^*_{u\!}\!\approx\,$338, $m^*_{d}\!\approx\,$322, $m^*_{s}\!\approx\,$510 (in MeV units). 
One can estimate the magnetic moment of $c$ quark to be $\mu_c$=+0.4$\mu_N$ (a significant value) 
using $m^*_c \approx 1.5^{\,}$GeV.  

However, for virtual $(q\bar q)$ quarks, the constituent masses are unjustifiably large
and one should use smaller (e.g. {\it current\,} quark \cite{PDG}) masses: $\tilde m_u$=$^{\,}2.3$, $\tilde m_d$=4.8, $\tilde m_s$=$^{\,}95$
 and
$\tilde m_c$=$1270^{\,}$(MeV). In such a case,  magnetic moments $\tilde \mu_q$ of light virtual quarks are scaled up 
according to $ (m^*_q/\tilde m_q)$ ratios, which gives factors $135\times $,  $67\times $, and $5\times $  for $u, d, s$ quarks.
The {\it current\,} quark magnetic moments thus are: $\tilde\mu_u$=+$250\mu_N$, $\tilde\mu_d$=$-65\mu_N$,
 $\tilde\mu_s$=$-3\mu_N$  (and $\tilde\mu_c$=+$0.47\mu_N$).

Using $\tilde\mu_{\bar s\,}$=$^{\,}3\mu_N$ for {\it current} quark $\bar s$, 
one has $|\tilde\mu_{s} B| \approx 1^{\,}$MeV
for $B\approx 10^{13\,}$T in the nucleon ($\mu_N$=$3.15\times 10^{-14\,}$MeV/T). 
Assuming hyperfine splitting $\Delta E^{s\bar s}_{10} \approx 100^{\,}$MeV, 
for $^{3\!}\tilde P_{00}$ and $^{1\!}\tilde P_{10}$ states
of virtual $(s\bar s)$ pair, mixing parameter $\varepsilon_B = 2\tilde\mu_s B/\Delta E^{s\bar s}_{10} = 0.019$. 
From Eq.(\ref{MixingB03PoSpin}) we obtain
\begin{equation}
\Delta s = s\!\uparrow\! - s\!\downarrow \,= [(1-\varepsilon_B)^2 -  (1+\varepsilon_B)^2]/6 = -0.0126\quad ; \qquad 
\Delta \bar s = \bar s\!\uparrow\! - \bar s\!\downarrow \, = +0.0126 \,\,
\end{equation}
for proton, and $(\Delta s $+$\Delta \bar s)$=$0$. For virtual $c$ quarks we have
$\Delta c \approx +0.002$, using $|\tilde \mu_s/\tilde\mu_c| \approx 6.4$, $\Delta E^{c\bar c}_{10} \approx 100^{\,}$MeV.
One can also expect $\Delta \bar d > \Delta \bar s >0$ and $\Delta \bar u < 0$,
for the virtual anti-quarks.

Parallel orientation of magnetic moments $\uparrow_{\!\mu_s\!}\uparrow_{\!\mu_{\bar s}}$ is more 
probable than the opposite $(\downarrow^{\!\mu_s\!}\downarrow^{\!\mu_{\bar s}})$ one in the polarized $^{3\!}\tilde P_{00}$ state. 
Induced magnetic moment $\langle \mu\rangle_{\!s\bar s}$ of virtual $(s\bar s)$ pairs 
can thus contribute \cite{Nature} to the magnetic moment of proton, while the overall spin as well as
total (and orbital) angular momentum of virtual ($s\bar s$) pairs  
remain to be zero. 

\section{Summary and conclusions}
We have considered a simple model of $^{3\!}P_{00} [q\bar q]$ quantum state polarization of virtual 
$(s\bar s)$ and $(c\bar c)$ pairs in the nucleon, occuring due to the internal magnetic field $B\approx 10^{13\,}$T, 
which originates from the nucleon magnetic moment. 
Our estimates are based on the analogy with the behavior of $^{3\!}P_{00}$ state of Positronium ($e^+e^-$) in the 
magnetic field \cite{Positr16x16}. 
The angular momentum 
of virtual $(s\bar s)$ pairs remains zero, while
$s$ quarks are polarized oppositely (and $\bar s$ parallel) to the proton spin.
The estimated polarizations are: 
$- \Delta \bar s  = \Delta s \approx - 0.013$, and $\Delta c = - \Delta \bar c \approx  +0.002$.

However, quarks interact also via strong interactions, and QCD effects \cite{AlbergPLB} can be 
larger than the electromagnetic interactions we have considered here. Moreover, internal volume of the
polarized nucleon presumably contains an oriented chromo-magnetic fields $B^a$ \cite{BA}. The interaction 
 $H^{int}_{\scriptscriptstyle QCD} = - \mu^a_q B^a$
of such gluonic fields $B^a$ with chromo-magnetic moments $\mu^a_q$ of virtual quarks 
 may induce a stronger polarization of $^{3\!}P_{00} [q\bar q]$ state. 
Our inferences about the $(s\bar s)$ polarization magnitude, 
the polarization orientation of $(c\bar c)$ pairs, 
and the opposite polarization of $\bar s,\bar d$, and $\bar u$ quarks in neutron (relative to proton case) should thus be taken with caution.

\ack Author is indebted to the organizers of DSPIN 2017 conference for the invitation to participate in this valuable
scientific meeting, and to JINR, for the financial support and kind hospitality.

\newpage
\end{document}